\documentstyle[epsf]{article}
\newcommand{\be}{\begin{eqnarray}}
\newcommand{\ee}{\end{eqnarray}}

{
{

\def\0n{0\nu\beta\beta}

\begin{document}
\thispagestyle{empty}
\begin{center}
{\Large \bf Implications of observed 
neutrinoless double beta decay} \\
\vspace{1.5cm}
{\large H.V. Klapdor-Kleingrothaus $^{\dagger}$
and U. Sarkar$^{\dagger,\ddagger}$}\\
\vspace{0.75cm}
{$^{\dagger}$ Max-Planck-Institut f\"ur Kernphysik,
P.O. 10 39 80, D-69029 Heidelberg, Germany} \\ \vspace{0.3cm}
{$^{\ddagger}$ Physics Department, Visva-Bharati University,
Santiniketan 731 235, India}
\end{center}

\vspace{2.5cm}

\begin{abstract}

Recently a positive indication of the neutrinoless double
beta decay has been announced. 
We study the implications of this result
taking into consideration earlier results on atmospheric
neutrinos and solar neutrinos. We also include in our
discussions the recent results from SNO and K2K. 
We point out that on the confidence level given for the double
beta signal, the neutrino mass matrices are now highly 
constrained. All models predicting Dirac masses are ruled
out and leptogenesis becomes a natural choice. 
Only the degenerate and the inverted hierarchical 
solutions are allowed for the three generation 
Majorana neutrinos. In
both these cases we find that the radiative corrections
destabilize the solutions and the LOW, VO and Just So solutions
of the solar neutrinos are ruled out. For the four generation
case only the inverted hierarchical scenario is allowed.

\end{abstract}

\vspace{1cm}
\newpage

Recent evidence of non-zero neutrino mass in the atmospheric 
neutrino anomaly \cite{atm},
strong constraints from the solar neutrinos \cite{sol}
and some positive indications from Laboratory experiments
\cite{k2k,lsnd} have already
restricted the possible neutrino mass matrices to only
a few possible choices \cite{alt}. 
The atmospheric neutrino result is consistent with
an oscillation between the $\nu_\mu$ and $\nu_\tau$ 
with maximal mixing.\footnote{Even though the $L/E$ flatness 
of the electron-like event ratio in the full three-flavor
framework suggests a bi-maximal mixing \cite{dharam}, here
we shall restrict to the simplest scenario in which one explains
the atmospheric neutrino anomaly via    
$\nu_\mu \leftrightarrow \nu_\tau$ oscillations. In addition, 
in our analysis we shall not incorporate the quantum gravity 
effects that may induce additional modifications to neutrino 
oscillations  \cite{dharam1}.}  
There are several allowed regions in the parameter space
of the mass squared differences and mixing angles of the
neutrinos, which can explain the solar neutrino problem
\cite{atman,solan}. But none of these
results could tell us about the absolute value of the
neutrino mass and whether the neutrinos are Majorana
or Dirac particles, {\it i.e.}, whether the neutrino mass
conserves lepton number or not. 

A positive signal of the neutrinoless double beta decay
has recently been announced \cite{ndbex}. The $0 \nu \beta
\beta$ decay describes a process in which two electrons 
are released (with lepton number two)
without any associated antineutrinos, and hence
lepton number is broken \cite{ndbook}. 
\begin{figure}[thb]
\vskip 0in
\epsfxsize=50mm
\centerline{\epsfbox{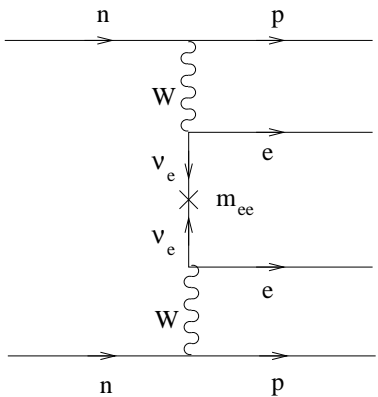}}
\vskip 0in
\caption{
Diagram mediating neutrinoless double beta decay. 
\label{summ1}}
\end{figure}
This process can be
mediated by a light virtual electron neutrino $\nu_e$ 
(as shown in figure \ref{summ1}), whose effective Majorana mass 
$$ {\cal L}_{Maj} = m_{ee} \nu_e \nu_e $$
breaks the lepton number. This physical electron neutrino 
state $\nu_e$ is the neutrino which couples to the
physical electron. So, $m_{ee}$ is the 
$(11)$ element of the neutrino mass matrix $M^\nu_{\alpha 
\beta}$, ($\alpha = e, \mu, \tau$)
in the basis in which the charged lepton mass matrix
is diagonal. The physical state $|\nu_\alpha>$ is related to
the mass eigenstates $ \nu_i$ (with eigenvalues
$m_i$, $i=1,2,3$) by the mixing matrix $U_{\alpha i}$ by
\begin{equation} |\nu_\alpha > = U_{\alpha i} |\nu_i > 
\label{mix}
\end{equation}
so that the mass matrix in the flavour basis
is related to the diagonal mass matrix through
\begin{equation}
M^\nu_{\alpha \beta} = U_{\alpha i} M_{ij}^{\rm diag} 
U^T_{\beta j} 
\end{equation}
where $M_{ij}^{\rm diag} = m_i \delta_{ij}$. 
The $0 \nu \beta
\beta$ decay rate depends on this effective Majorana mass
\cite{maj,wolf}
\begin{equation}
m_{ee} = M^\nu_{ee} = \sum_i |U_{e i}|^2 m_i .
\label{mix1}
\end{equation}
Taking care of the nuclear
matrix elements and other factors, the present signal for
the $0 \nu \beta \beta$ decay amounts to an effective Majorana
mass of the electron neutrino in the range of
\begin{equation}
m_{ee} = (0.05 - 0.86) ~ {\rm eV} ~~~~~~~ {\rm at~ 95\%~ ~c.l.}
\end{equation}
with a best value of $0.4$ eV. This is the first indication of 
lepton number violation in nature and
the fact that Majorana fermions can exist
in nature. This result establishes that a fermion can
be its own antiparticle, violating lepton number by two
units. There are several
other consequences of this new result for physics
beyond the standard model \cite{cons}. Here we study
some of the implications for the neutrino masses.

In the standard model neutrinos are massless. One may
extend the model with one singlet 
right-handed neutrino $N_{aR}$ per 
generation. Then the interaction $h_{ia} \bar 
\ell_{iL} N_{aR} \phi$ can give a Dirac mass, where $h_{ia}$
are the Yukawa couplings, $\ell_{iL}$ are left-handed
leptons and $\phi$ is the usual Higgs doublet.
For the neutrino mass to be of the order of eV, the Yukawa
couplings have to be extremely small, $h \sim 10^{-12}$. 
Although such small numbers are not natural, this is surely
a possibility. If this is the only source of neutrino mass,
then the contribution to the neutrinoless double beta decay
will come from two diagrams, one from an exchange of $\nu_e$
and the other from an exchange of $N_R$. These two diagrams
will cancel each other exactly \cite{wolf}. A naive way of
understanding this result is to consider the lepton number.
Since $\ell_{iL}$ and $N_{aR}$ both carry lepton number one,
the above Dirac mass term does not violate lepton number.
So this term cannot allow a lepton number violating process
like $0 \nu \beta \beta$ decay.
Thus the new result rules out this unnatural possibility 
of Dirac neutrinos with unnaturally tiny Yukawa couplings
completely.

A natural choice for small neutrino mass is to consider
an effective five dimensional operator \cite{path}
with only the left-handed leptons 
\begin{equation}
{\cal L}_{eff} = {f_{ij} \over M} \ell_{iL} \ell_{jL}
\phi \phi  .
\end{equation}
For any large lepton number violating scale $M$, this
operator would then give a very small Majorana mass to
the neutrinos. Within the context of the standard model
this is the only effective term allowed, which can give 
a neutrino mass. However, since this is not a 
renormalizable term, this term cannot originate from
the standard model alone. There has to be some higher
theory at the cut-off scale $M$, which is the lepton 
number violating scale in this case. So, this already
indicates the nature of physics beyond the standard
model. 

This operator has several possible realizations. 
In the see-saw mechanism \cite{seesaw}
one breaks lepton number with a Majorana mass
of the right-handed neutrinos
$$ {\cal L}_{N} = M_{Nab} N_{aR} N_{bR}  $$
The interplay of the Dirac mass term and this term then
induces a lepton number violating effective neutrino mass
term $$ {\cal L}_{\nu} = {h_{ai} h^T_{bj} < \phi >^2 
\over M_{Nab}} \nu_{iL} \nu_{jL} ,$$ which can give rise
to the neutrinoless double beta decay.

In another realization of the effective dimension five
operator a triplet Higgs is introduced \cite{trip1,trip}.
In one version of the model \cite{trip1}, lepton number
was broken with a vacuum expectation value ($vev$) of 
the triplet Higgs, resulting in a Majoron. These scenarios
are ruled out from the Z-width at LEP. A newer version
of the model now breaks lepton number explicitly at a 
very high scale $M \sim M_\xi \sim \mu$ through the 
couplings of the triplet Higgs $\xi$ 
$$ {\cal L}_{\xi} = M_\xi^2 \xi^\dagger \xi +
f_{ij} \ell_{iL} \ell_{jL} \xi
+ \mu \xi^\dagger \phi \phi .$$
The minimization of the complete potential for $\mu \neq 0$
then gives a small $vev$ to the triplet Higgs 
$<\xi> \sim {\mu <\phi>^2 \over M^2_\xi}$, which
generates a Majorana mass of the neutrino. 

If one extends the standard model to a larger left-right
symmetric model, then in the left-right symmetric model
neutrinos can acquire mass from both the see-saw mechanism
and the triplet Higgs mechanism \cite{lr}. Then there are
radiative mechanisms \cite{zee}
with a low lepton number violating scale. 

If we now consider the dimension five operator,
the effective Majorana neutrino mass is $m_{ee} \sim
f_{11} <\phi>^2/M$. For $f_{11} \sim 0.01$ the present 
result of $0 \nu \beta \beta$ decay implies a lepton
number violating scale of $M \sim 10^{10}$ GeV.
This lepton number
violating scale can then address another important 
question of the baryon asymmetry of the universe
\cite{lepto1,lepto2}.
In both the see-saw \cite{lepto1} and the triplet 
Higgs models \cite{trip}, 
with lepton number violating scale it is possible
to generate a lepton asymmetry of the universe. 

In the see-saw mechanism, when the right-handed neutrinos
decay into leptons and antileptons, letpon number is 
violated at the temperature $T=M_{Nab}$. 
The Majorana phases of the right-handed
neutrinos give enough CP violation in these decays and
the out-of-equilibrium condition is naturally satisfied
at this scale. This would then generate a lepton asymmetry
of the universe. In the triplet Higgs models the decays of
the triplet Higgs to Higgs doublets and the leptons
violate lepton number at the temperature $T=M_\xi$. 
The coupling of the triplet Higgs, $f_{ij}$ and $\mu$,
contains the required CP violation. 
Before the electroweak phase transition this
lepton asymmetry would then get converted to the
baryon asymmetry of the universe in the presence of
the sphalerons \cite{sph}. 

Recently a new possibility of TeV scale gravity with
extra dimensions has become very promising phenomenologically
\cite{extra}. In these models all the standard model
particles are confined only in our 4-dimensional world,
while gravity can propagate in all the directions,
including the extra dimensions forming most of the bulk
of space-time. Although gravity is strong in the extra 
dimensions, the small overlap of our world with the extra
dimensions makes gravity weak in our world. In these
models of extra dimensions one includes a right-handed 
neutrino field in the bulk \cite{exnu1}.
The overlap of the right-handed neutrino in our world
will then be small and this can make the Yukawa coupling
$h_{ia}$ in $h_{ia} \bar \ell_{iL} N_{aR} \phi$ naturally 
small, so that now there can be a very small Dirac mass
of the neutrinos. The Yukawa coupling $h_{ia}$ is 
naturally suppressed by the volume of the extra dimensions.
There are several new phenomena associated with 
these scenarios including new predictions
for neutrino oscillations \cite{exnu1}. All these models are now 
ruled out by the new observation. 

The $ 0 \nu \beta \beta$ decay would now allow only a
few possible choices for the neutrino mass in these theories
of large extra dimensions \cite{extrip,exmaj}. 
In one possibility \cite{extrip} lepton number is
broken in a distant brane and the effect shines in our
world to give a small lepton number violating coupling 
of a Triplet Higgs scalar, which then generates a small
neutrino mass. In this scenario the triplet Higgs will
have definite same sign dilepton signals in the next
generation colliders, which should be observed \cite{extrip}.
Of course one may consider an effective operator in the
higher dimensions and can consider a Majorana mass, or
break lepton number in the bulk to generate Majorana
masses \cite{exmaj}, 
details of realization of such models are yet to be studied. 

Given the model of generating a small Majorana mass, 
the exact structure of the mass matrix may be such that there
are two contributions from different mass eigenstates, 
which cancel each other forbidding $ 0 \nu \beta \beta$ decay.
Thus the present observation of the $ 0 \nu \beta \beta$ decay
would rule out a class of models where this cancellation
takes place partially or fully. Moreover constraints from
the atmospheric and solar neutrinos and Lab limits could
be used in conjunction with the present limit on 
$ 0 \nu \beta \beta$ decay to discriminate some of the
possible mass spectra following the analysis of 
\cite{smir} and with one new consideration that 
in some cases electroweak radiative corrections destabilize
the LOW, VO and Just So solutions of the solar neutrinos.

Taking the latest analysis of the solar neutrino data,
including SNO, eight solutions are allowed \cite{solan}.
For three generation of neutrinos the solutions are 
categorised as (LMA) large
mixing angle MSW solution; (SMA) small mixing angle MSW
solution; (LOW) low probability low mass solution; (VO) 
vacuum oscillation solutions; (Just So) very low mass squared
difference vacuum oscillation solutions.
For our analysis, the three 
solutions LOW, VO and Just So, will not make much
difference, so we call them together VAC solution.
With sterile neutrinos only the SMA, VO and Just So solutions
are allowed. The atmospheric neutrino anomaly and the
initial results from K2K determine the $\nu_\mu \to \nu_\tau$
mixing and mass squared difference \cite{atm,k2k}. These results are
summarized in table \ref{numa}. Several of these solutions
will now be ruled out by the new neutrinoless double beta decay
result as mentioned in table \ref{numa}.

\begin{table}
\caption{Allowed neutrino 
mass squared differences ($\Delta m^2$) and mixing angles
($\sin^2 2 \theta$) from solar and atmospheric neutrino
results according to \cite{solan,sol,atman,atm,k2k}. 
All masses are in eV. The last column shows 
if any particular solution is allowed by the new result
from $ 0 \nu \beta \beta$ decay and our present analysis. }
\begin{center}
\begin{tabular}{lccc}
\hline \hline
Solution & $\Delta m^2$& $\sin^2 2 \theta$& 
$[(\beta \beta)_{0 \nu}]$ \\
\hline
&&&\\
\multicolumn{4}{c}{Three generation Solar neutrino
solutions} \\
&&&\\
SMA & $(4 - 9) \times 10^{-6}$ & $(.0008 - .008) 
$&Allowed \\
LMA & $(2 - 20) \times 10^{-5}$ & $(0.3-0.93)
$& Allowed\\
LOW & $(6 - 20) \times 10^{-8}$ & $(0.89 - 1)
$& Not Allowed\\
VO & $10^{-10}$ & $(0.7-0.95)
$& Not allowed\\
Just So & $(5 - 8) \times 10^{-12}$ & $(0.89 - 1)
$& Not allowed\\
&&&\\
\multicolumn{4}{c}{Solar neutrino
solutions with sterile neutrino} \\
&&&\\
SMA & $(3 - 8) \times 10^{-6}$ & $(.0006 - .008)
$& Allowed\\
VO & $1.4 \times 10^{-10}$ & $(0.7 - 0.9)
$& Allowed\\
Just So & $(6 - 8) \times 10^{-12}$ & $(0.82 - 1)
$& Allowed\\
&&&\\
\multicolumn{4}{c}{Atmospheric neutrino} \\
&&&\\
$\nu_\mu \to \nu_\tau$ & 
$(1.8 - 4.0) \times 10^{-3}$ & $(0.87 - 1.0)$&
Consistent\\
&&&\\
\hline \hline
\end{tabular}
\label{numa}
\end{center}
\end{table}

First consider only three generations of
neutrinos. The mixing matrix of equation [\ref{mix}]
may be parametrized as
\begin{equation}
U = \pmatrix{ c_1 c_3 & - s_1 c_3 & -s_3 \cr
s_1 c_2 - c_1 s_2 s_3 & c_1 c_2 + s_1 s_2 s_3 &
- c_3 s_2 \cr s_1 s_2 + c_1 c_2 s_3 &
c_1 s_2 - s_2 c_2 s_3 & c_2 c_3 } ,
\end{equation}
where, $s_i = \sin \theta_i$ and $c_i = \cos \theta_i$.
$s_2$ represents $\nu_\mu - \nu_\tau$ mixing and is
determined by the atmospheric neutrino anomaly, $s_3$ gives
$\nu_e - \nu_\tau$ mixing and is constrained by the
CHOOZ result and $s_1$ is related to the solar neutrinos.
Then the contribution to the
$ 0 \nu \beta \beta$ decay is given by
\begin{equation}
m_{ee} = \sum_i |U_{e i}|^2 m_i =
m_1 ~c_1^2 c_3^2 + m_2 ~ s_1^2 c_3^2 + m_3 ~ s_3^2  .
\label{ndb}
\end{equation}
$m_i$ are complex and contain the Majorana phases,
which can contribute to the lepton number violating
$ 0 \nu \beta \beta$ decay.
The phase in the mixing matrix $U$ may only show
up in the neutrino oscillation experiments and does not 
contribute to the lepton number violating 
$ 0 \nu \beta \beta$ decay. 

The contributions of the different mass patterns 
to the $ 0 \nu \beta \beta$ decay may now be estimated
for the allowed 
values of the mixing angles ($s_2$ and $s_3$)
and the mass eigenvalues ($m_1$, $m_2$ and $m_3$).
We allow possible variation of the phases in the
masses to check for cancellation. This gives a range 
for $m_{ee}$, which is similar to the values obtained
in an earlier analysis \cite{smir}. 
There are some changes, which
come due to the new input from SNO. Since the mixing
angle $s_2$ does not enter the expression 
[equation (\ref{ndb})], new results from K2K do not
affect the analysis. In addition we consider 
electroweak radiative corrections and point out that
they may destabilize the very small mass differences
required for some of the solar neutrino solutions. 

The different three generation models of the
neutrino masses may be classified as: 
\begin{quote}
{\bf Hierarchical} The masses satisfy a hierarchical pattern
$$ m_1 \ll m_2 \ll m_3 , $$ $${\rm so~ that}~~~ m_3 = m_{atm} =
\sqrt{\Delta m_{atm}^2} ~~~~ {\rm and} ~~~~ m_2 = m_{sol}
= \sqrt{\Delta m_{sol}^2} . $$
{\bf Degenerate} All the three masses are of the same order
$$ m_1 \approx m_2 \approx m_3 = m_0 .$$
Atmospheric neutrinos require $m_0 > 0.042$ eV, hot dark matter
prefers $m_0 \sim $eV, but the neutrinoless double beta decay imposes
$m_0 < 0.8$ eV. The mass squared differences are much smaller
$$ \Delta m_{12}^2 = \Delta m_{sol}^2 ~~~~~~ {\rm and} ~~~~~~
\Delta m_{23}^2 = \Delta m_{atm}^2 $$
{\bf Partially Degenerate} The lighter masses are almost 
degenerate and their mass squared difference explains solar
neutrinos 
$$ m_1 \approx m_2 < m_3 \approx m_{atm}
= \sqrt{\Delta m_{atm}^2}  ,$$
with $ \Delta m_{12} \approx \Delta m_{sol} .$ \\
{\bf Inverted Hierarchical} $\nu_e$ is part
of the heavier states and the mass splitting between $m_2$ and
$m_3$ explains the solar neutrino puzzle
$$ m_1^2 \ll m_3^2 \approx m_2^2 = \Delta m_{atm}^2
~~~~ {\rm and} ~~~~ \Delta m_{23}^2 \approx \Delta m_{sol}^2 .$$
\end{quote}

\begin{figure}[thb]
\vskip -4in
\epsfxsize=250mm
\centerline{\epsfbox{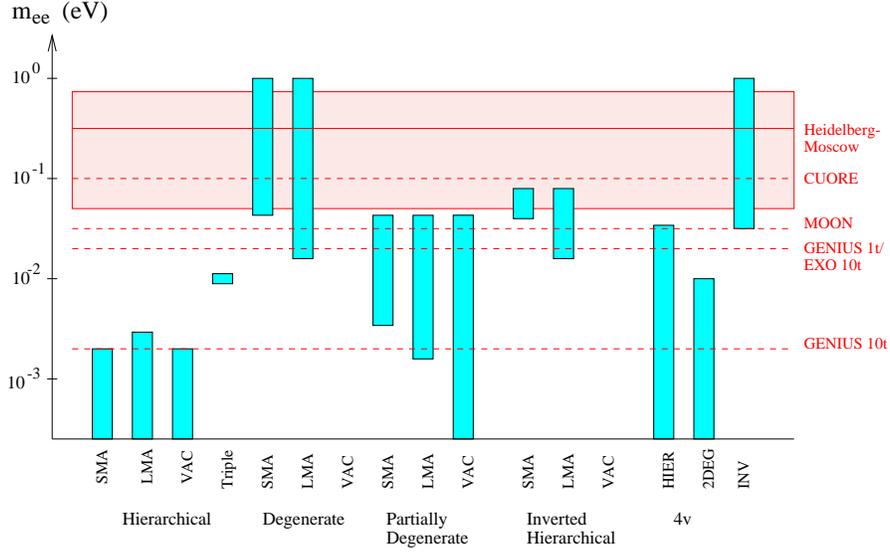}}
\vskip 1.6in
\caption{
Contributions in different models to the
neutrinoless double beta decay.
The present result is given by the shaded region
(the solid line denoting the best value),
which allows only few of the models. The 
expected sensitivity for the CUORE \cite{cuore},
MOON \cite{moon} and the one ton and ten tons GENIUS
\cite{genius} are given for comparison.
\label{summ}}
\end{figure}

In the hierarchical scenario for SMA solution
the largest contribution to $ 0 \nu \beta \beta$ decay 
comes from $m_3$ inspite of the small mixing angle $s_3$ 
constrained by CHOOZ, so that $m_3 s_3^2 < 0.002$ eV.
For the LMA solution there is another contribution from
$m_2 \sim m_{sol} \sim 0.005$, which is comparable 
and bounded by $0.0004 < m_2 \sin^2 \theta < 0.0015$.
Since there is no lower bound on $m_3$ contribution,
both these contributions can cancel each other and again
there is no lower bound. The upper bound is now $m_{ee} <
0.0035$. Thus the contribution to the
$ 0 \nu \beta \beta$ decay is much smaller 
than that allowed by the present observation for
all the hierarchical solutions to the solar neutrino problem.
Another possibility (triple) with hierarchical mass matrix
is to consider all elements of the mixing matrix to be
equal, which has a definite prediction for the $0 \nu 
\beta \beta$ decay $m_{ee} \sim 0.02$ and is also ruled out. 
For the different solutions of the hierarchical scenario,
the predictions differ slightly, but all of them
fall below the allowed region. This is shown in
the summary figure \ref{summ}. 

For partially degenerate neutrino masses,
$ 0.005 ~{\rm eV} < m_1 < 0.042 ~{\rm eV} $
restricts the amount of $ 0 \nu \beta \beta$ decay.
For SMA the main contribution comes from $m_1$ and
hence $m_{ee}$ is bounded by the value of $m_1$.
For the LMA solar neutrino solution the contribution
from $m_2$ is comparable but smaller, so considering
the bound on the mixing angle, the bound becomes
$0.042 > m_{ee} > 0.0015$. For the VAC solutions the
mixing angle could be maximal and the contributions
from $m_1$ and $m_2$ can cancel, so there is no lower
bound. In all the cases the prediction for
$ 0 \nu \beta \beta$ decay comes out to be
smaller than the presently acceptable
range for all the solutions of the solar neutrinos
and hence all these solutions are ruled out. 
In ref. \cite{ndbex} the authors 
included the transition region (which is $m_1 \approx
m_2 \approx m_3$) in their definition of the
partial degenerate solution. As a result they find
an overlap of the partial degenerate solution with
the present result of $0 \nu \beta \beta$ decay. 
But we have included the transition region in the
degenerate mass spectrum and defined out partially
degenerate solution as $m_1 < m_3$, so that the
upper bound on $m_1$ gives the bound $m_{ee} < m_3 
= m_{atm} = 0.042$.

The degenerate mass spectrum can be realized, simultaneously 
providing a solution to the solar and atmospheric
neutrinos, with $0.042 ~{\rm eV} \leq m_1 \leq 1$ eV. 
In this case the contribution
to $m_{ee}$ comes mostly comes from $m_1$ for the SMA. But for
the LMA both $m_1$ and $m_2$ contribute and there is also
a possibility of partial cancellation. Considering $s_1 
< 0.93$ the lower bound comes out to be 
$m_1 c_1^2 - m_2 s_2^2 > 0.015$. Both the
SMA and the LMA solutions are thus allowed by the
present observation of $ 0 \nu \beta \beta$ decay.
But the VAC solutions for solar neutrinos,
which require $m_1^2 - m_2^2 < 10^{-8}$ eV, are ruled out
by the following argument.

In the basis in which the charged lepton masses are diagonal,
the diagonal terms in the neutrino mass matrices $M_\nu$ will 
receive a contribution from higher order weak interaction
corrections \cite{wolf1}. Consider for simplicity a mass matrix
in the basis $[\nu_e ~~~ \nu_\tau]$
$$ M_\nu = \pmatrix{ a & b \cr b & c} .$$
The degeneracy of the eigenvalues requires (1) $a = c = 0$ or 
(2) $a = -c = m_0$. In case 1, the radiative correction does not 
change the eigenvalues and we have a Dirac neutrino. But
in case 2, the radiative corrections break the degeneracy and
give us a pseudo-Dirac neutrino \cite{wolf1}. 
The present result on $0 \nu \beta \beta$ decay 
requires $a = m_{ee} \neq 0$, so we are
forced to consider case 2. In this case there is a
mass splitting between the two degenerate states, given
by the mass squared difference
\begin{equation}
\left( m_2^2 - m_1^2 \right)_{rad} \sim \alpha_W m_0^2 \left( 
{m_\tau^2 - m_e^2 \over m_W^2} \right) \sim
1.7 \times 10^{-5} ~m_0^2 ~ {\rm eV}^2 .
\end{equation}
$\alpha_W$ is the weak fine structure constant.

To explain the
atmospheric neutrino anomaly, the degenerate mass
should be $m_0 > 0.042$ eV. Then the radiative corrections
modify the tree level mass squared difference by 
$ \left( m_2^2 - m_1^2 \right)_{rad} > 1.5 \times 
10^{-8} ~ {\rm eV}^2 . $ With this correction it will
not be possible to maintain the mass squared difference
required for the VAC solutions and hence these VAC solutions
are ruled out in the degenerate mass scenarios. Although
we demonstrated with two generation example, this result
is applicable to three generation case. When we discuss
the specific textures we shall prove this generality.

This is true for the inverted hierarchy scenario as well.
In this case the radiative correction comes out to be the
same as above, with $m_0 = m_2 \approx m_3$. Thus the 
radiative correction would give a mass squared difference,
larger than acceptable by the VAC solutions. 
So, even if there is a model to predict
such small mass squared difference required by the VAC
solutions, after including the weak radiative correction
the model cannot maintain the required mass 
squared difference.

To demonstrate this result let us now work with one example.
We assume a mass matrix in the basis with diagonal charged leptons
\begin{equation}
M_\nu = m_0 \pmatrix{ 1 & 0 & 0 \cr 0 & 1/2 & 1/2 \cr 0
& 1/2 & 1/2} .
\end{equation}
which can be diagonalised to
$$ M^{diag} = {\rm Diag}~
\{ m_3,m_2, m_1 \} = m_0 ~ {\rm Diag} ~ \{ 1,~ 1,~ 0 \} .$$
This predicts maximal mixing between $\nu_\mu$ and
$\nu_\tau$. 
Inclusion of the radiative corrections would change
this mass matrix to 
\begin{equation}
M_\nu = m_0 \pmatrix{1+ \epsilon_e & 0 & 0 \cr 0 & (1+
\epsilon_\mu) /2 & 1/2 \cr 0 & 1/2 & (1+ \epsilon_\tau) /2} 
\end{equation}
where $\epsilon_i = \alpha_W ( m_i^2 / m_W^2 )$.
After diagonalisation the new mass eigenvalues will
become $m^{diag}_\nu \sim {\rm Diag}~
m_0 ~\{ 1+\epsilon_e, 1+\epsilon_\mu +\epsilon_\tau , 
\epsilon_\mu + \epsilon_\tau \}$, so that the
mass squared difference between the two degenerate
states has become $m_2^2 - m_3^2 \sim
1.7 \times 10^{-5} ~m_0^2 ~ {\rm eV}^2 $. For 
$m_0 \sim m_{atm} > 0.042$ this is 
too large for the VAC solutions. 

In this inverted hierarchical case the LMA and SMA
solutions of solar neutrinos are still allowed by
the present result of the $ 0 \nu \beta \beta$ decay.
Since the heavier states $\nu_2$ and $\nu_3$ contain
the $\nu_e$, they contribute to the $ 0 \nu \beta \beta$ decay.
But the mass is now restricted by the solution to the
atmospheric neutrino anomaly, since $m_2 \sim m_{atm}$,
we get $0.063 > m_{ee} > 0.042$. 
For the LMA solution for solar
neutrinos there can be cancellation but with the present
limit \cite{solan} on mixing for the LMA solution 
the cancellation can only be partial and the bound
is $0.063 > m_{ee} > 0.015$. 

Thus for the three generations of neutrinos, the LOW,
VO and the Just So solutions of the solar neutrinos are
not allowed. The hierarchical and partially degenerate
mass spectrum for all solutions of the solar neutrinos
are ruled out. Only the degenerate and inverted 
hierarchical mass spectrum are allowed which can provide
SMA and LMA solutions for solar neutrinos. There are some
theoretical problems with these cases. If one starts from
a grand unified theory and tries to evolve the Yukawa couplings
in supersymmetric models, it becomes difficult to 
maintain the degeneracy of two or three states
\cite{lola}. It
has been pointed out that if there is some symmetry
which protects some of the texture zeroes in the mass 
matrix, then these zeroes are protected against the 
renormalization group evolution \cite{prot}. There could be another
possibility to this problem of protecting the degeneracy. In
models of large extra dimensions the evolution of the 
Yukawa couplings can allow degenerate solutions. 

Considering this model building point of view, it is
convenient to study some textures of the neutrino mass
matrix. Earlier they were
considered to explain the atmospheric neutrino problem
assuming that the mass squared difference required
by the solar neutrino would come as
perturbation to these texture mass matrices. Since
the solar neutrino requires small mass squared difference,
this assumption is justified. 

In the degenerate and the inverted hierarchical scenarios,
which are now allowed by the $0 \nu \beta \beta$ decay
result, 
all the zeroth order neutrino mass matrices ($M_\nu$) with 
texture zeroes can be listed \cite{alt}

\vbox{
\begin{eqnarray}
M_\nu^{B1} &=& m_0 ~\pmatrix{1 & 0 & 0 \cr 0 & -1/2 & -1/2 \cr
0 & -1/2 & -1/2 } \hskip .5in M_\nu^{\rm diag} 
= {\rm Diag}~\{ m_0, -m_0, 0 \}
\nonumber \\
M_\nu^{B2} &=& m_0 ~\pmatrix{1 & 0 & 0 \cr 0 & 1/2 & 1/2 \cr
0 & 1/2 & 1/2 } \hskip .5in \phantom{--}
M_\nu^{\rm diag} = {\rm Diag}~\{ m_0, m_0, 0 \}
\nonumber \\
M_\nu^{C0} &=& m_0 ~\pmatrix{1 & 0 & 0 \cr 0 & 1 & 0 \cr
0 & 0 & 1 } \hskip .5in \phantom{--2/2}
M_\nu^{\rm diag} = {\rm Diag}~\{ m_0, m_0, m_0 \}
\nonumber \\
M_\nu^{C1} &=& m_0 ~\pmatrix{-1 & 0 & 0 \cr 0 & 1 & 0 \cr
0 & 0 & 1 } \hskip .5in \phantom{-/2/2}
M_\nu^{\rm diag} = {\rm Diag}~\{- m_0, m_0, m_0 \}
\nonumber \\
M_\nu^{C2} &=& m_0 ~\pmatrix{1 & 0 & 0 \cr 0 & 0 & -1 \cr
0 & -1 & 0 } \hskip .5in \phantom{/2/2}
M_\nu^{\rm diag} = {\rm Diag}~\{ m_0, -m_0, m_0 \}
\nonumber \\
M_\nu^{C3} &=& m_0 ~\pmatrix{1 & 0 & 0 \cr 0 & 0 & 1 \cr
0 & 1 & 0 } \hskip .5in \phantom{--2/2}
M_\nu^{\rm diag} = {\rm Diag}~\{ m_0, m_0, -m_0 \}
\nonumber 
\end{eqnarray}}

\noindent
where the $B$ solutions are for inverted hierarchical scenario
and $C$ solutions are for degenerate cases. For $B$ solutions,
$m_0^2 = \Delta m_{atm}^2$ and for $C$ solutions 
$m_0^2 > \Delta m_{atm}^2$. In case $C2$ and $C3$ it may appear 
that the radiative corrections may be small. But in case $C2$
it is not possible to have a solution to solar neutrinos with
maximal mixing, so the vacuum solutions are not possible. This 
mass matrix allows only a small mixing angle solution.

Let us now discuss case $C3$ in little details, which will
demonstrate how the radiative correction destabilize the 
vacuum solutions for three generation case. Let us say
$m_0 \gg m_{atm}$. To generate a mass difference $a \sim O(m_{atm})$
between $\nu_\mu$ and $\nu_\tau$ (with $m_{\nu_\mu}<
m_{\nu_\tau}$) without disturbing the degeneracy
between $\nu_e$ and $\nu_\tau$, we need to introduce a term
$a$ in the (33) element of the mass matrix. This will 
introduce a mass difference between $\nu_e$ and $\nu_\tau$
radiatively, which is $m_0~ a ~\epsilon_\tau > 10^{-8}$ eV, 
destabilizing
the VAC solutions in this case. Thus for all mass textures
and degenerate solutions, the VAC solutions are ruled out.

In all textures the mixing for 
the atmospheric neutrinos is considered to be 
maximal, $s_2 = c_2 = 1/\sqrt{2}$. It is also assumed
that the electron neutrino does not take part ($s_3=0$) in
atmospheric neutrino anomaly. Thus the mixing matrix can 
be parametrized only in terms of the solar neutrino 
mixing angle $s = s_1$ ($c = c_1$)
\begin{equation}
U = {1 \over \sqrt{2}} ~ \pmatrix{ \sqrt{2} c & - 
\sqrt{2} s & 0 \cr s & c & - 1 \cr s & c & 1} .
\end{equation}
For complete solutions suitable small perturbations need
to be added. But if at this level the $(11)$ element
vanishes, then with small perturbation it will not be
possible to predict the required amount of
$ 0 \nu \beta \beta$ decay. This criterion rules out several
of the mass matrices with texture zeroes \cite{alt}.

Let us now include sterile neutrinos in the discussion.
The latest result from Super-Kamiokande rules out
sterile neutrino mixing for only two generations \cite{skst}. In a
more general analysis with four generations it has been
shown that it is not ruled out, but the allowed parameter
space is restricted \cite{barg}. On the other hand the main motivation
of introducing the sterile neutrino is to explain the LSND
result \cite{lsnd}, which is in partial conflict with the 
KARMEN result \cite{karmen}.
Considering all this we shall not give much details with
sterile neutrinos. 

One possibile mass spectrum could be hierarchical, where
the sterile neutrino has a mass of the order of 
eV to explain LSND and
the other neutrinos have similar structure as the three
generation hierarchical case. The sterile neutrino mixes 
very weakly with other generations. So the contributions
to the $ 0 \nu \beta \beta$ decay coming from all the
states are small, and taking the maximum allowed values
for the different mixing angles this scenario predicts
$m_{ee} < 0.03$ eV. This possibility is thus 
not allowed by the present $ 0 \nu \beta \beta$ decay
result. Another possibility is that two of the heavier states
are composed of $\nu_\mu$ and $\nu_\tau$, whose mass 
difference explains the atmospheric neutrino problem. $\nu_e$ is
the lightest and its mixing with $\nu_\mu$ explains
LSND. This scenario is also ruled out, since it predicts
$m_{ee} < 0.01$.

The inverted hierarchical scenarios are not ruled out, where
$\nu_e$ and $\nu_s$ [or $\nu_\tau$] have mass to explain
LSND and small mass squared difference to explain solar
neutrinos. The reason is that the mass difference between $\nu_\mu$ 
and the fourth neutrino 
is such that they explain the atmospheric neutrino problem. For large
mixing there will be cancellation and the contribution
to the $ 0 \nu \beta \beta$ decay will be small. So
only a very restricted parameter space is now allowed by
the new $ 0 \nu \beta \beta$ decay result. 

In summary, we studied the implications of the first 
evidence for the neutrinoless double beta decay. This
is the first evidence for lepton number violation and
for a Majorana particle. So models of Dirac neutrinos 
are now ruled out and the models of Majorana neutrinos
can generate a lepton asymmetry of the universe. The
structure of the Majorana mass matrix also gets
more constrained. For the three generation case
the hierarchical and partially degenerate
neutrino mass matrices are not allowed. Only the degenerate
and inverted hierarchical models are allowed. Even 
in these two cases, if one considers electroweak radiative
corrections, the small mass solutions of the solar neutrinos
become difficult to accomodate. For the allowed scenarios
possible texture mass matrices are mentioned. In the 
four generation scenario, only the inverted hierarchical
models are allowed. 

\vskip .5in

\centerline{\bf Acknowledgement}

\vspace{0.5cm}
U.S. wants to thank the Max-Planck-Institut f\"ur Kernphysik,
Heidelberg, Germany for hospitality.

\end{document}